\documentclass[sigconf,nonacm]{acmart}
\AtBeginDocument{%
  }

\setcopyright{cc}
\copyrightyear{2025}
\acmYear{2025}
\acmDOI{XXXXXXX.XXXXXXX}

\newcommand{\benchmark}{\texttt{EnronQA} }
\usepackage[most]{tcolorbox}

\begin{document}

\title{EnronQA: Towards Personalized RAG over Private Documents}

\author{Michael J. Ryan}
\authornote{Work completed during Snowflake internship.  Email: michaeljryan@stanford.edu}
\orcid{0000-0002-4353-821X}
\affiliation{%
  \institution{Stanford University}
  \city{Palo Alto}
  \state{California}
  \country{USA}
}

\author{Chris Nivera}
\affiliation{%
  \institution{Snowflake}
  \city{San Mateo}
  \state{California}
  \country{USA}}

\author{Danmei Xu}
\affiliation{%
  \institution{Snowflake}
  \city{San Mateo}
  \state{California}
  \country{USA}
}

\author{Daniel Campos}
\affiliation{%
 \institution{Snowflake}
 \city{San Mateo}
 \state{California}
 \country{USA}}

\renewcommand{\shortauthors}{Ryan et al.}

\begin{abstract}
Retrieval Augmented Generation (RAG) has become one of the most popular methods for bringing knowledge-intensive context to large language models (LLM) because of its ability to bring local context at inference time without the cost or data leakage risks associated with fine-tuning. A clear separation of private information from the LLM training has made RAG the basis for many enterprise LLM workloads as it allows the company to augment LLM's understanding using customers' private documents. Despite its popularity for private documents in enterprise deployments, current RAG benchmarks for validating and optimizing RAG pipelines draw their corpora from public data such as Wikipedia or generic web pages and offer little to no personal context. Seeking to empower more personal and private RAG we release the \benchmark benchmark, a dataset of 103,638 emails with 528,304 question-answer pairs across 150 different user inboxes.  \benchmark enables better benchmarking of RAG pipelines over private data and allows for experimentation on the introduction of personalized retrieval settings over realistic data.  Finally, we use \benchmark to explore the tradeoff in memorization and retrieval when reasoning over private documents. \footnote{All data released on this Huggingface repo: \href{https://huggingface.co/datasets/MichaelR207/enron_qa_0922}{MichaelR207/enron\_qa\_0922}}
\end{abstract}

\begin{teaserfigure}
  \includegraphics[width=\textwidth]{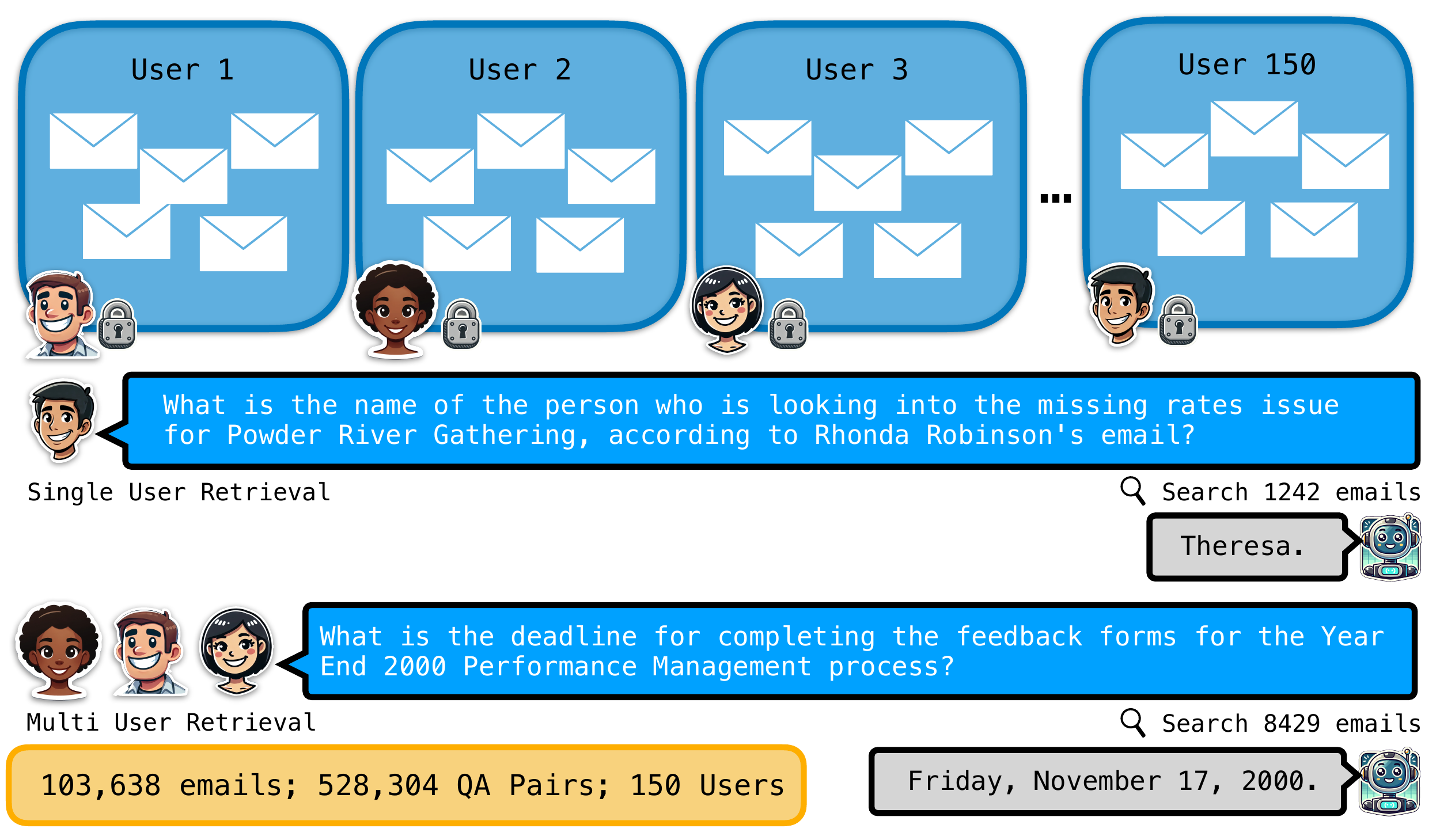}
  \caption{The \benchmark benchmark enables personalized and private retrieval benchmarking on a cleaned corpus of over 100,000 emails spanning 528,304 quality question-answer pairs over 150 users.  We explore both single and multi-user retrieval settings.}
  \Description{The \benchmark benchmark enables personalized and private retrieval benchmarking on a cleaned corpus of over 100,000 emails spanning \textcolor{red}{X} quality question-answer pairs over 150 users.  We explore both single and multi user retrieval settings.}
  \label{fig:teaser}
\end{teaserfigure}


\settopmatter{printccs=false, printacmref=false}
\maketitle

\section{Introduction}
Retrieval is increasingly one of the most common ways to add context to LLMs in a process called Retrieval Augmented Generation (RAG) \cite{grandview2024rag, lewis2021retrievalaugmentedgenerationknowledgeintensivenlp, gao2024retrievalaugmentedgenerationlargelanguage}.   RAG pipelines involve augmenting the natural language generation capabilities of an LLM with an external data store. This enhancement improves the factuality \cite{shuster2021retrieval, ayala-bechard-2024-reducing} and the explainability \cite{xia2024groundsentenceimprovingretrievalaugmented, 10.1145/3626772.3657660} of the LLM by grounding the generation in documents.  Furthermore, RAG has been shown to help LLMs solve knowledge-intensive tasks \cite{kandpal2023largelanguagemodelsstruggle} by retrieving relevant knowledge rather than relying on the LLM to memorize facts.  One of RAG's most popular applications is retrieval over private documents, enabling companies and users to interact with vast stores of internal and private knowledge \cite{10.1145/2661829.2661865, ragflow2024github}.

Although retrieval over private documents is one of RAG's most popular use cases, comparatively few large-scale RAG benchmarks focus on private document retrieval \cite{arora2023reasoning}.  Most popular benchmarks for validating and optimizing RAG pipelines draw their corpora from Wikipedia \cite{yang-etal-2018-hotpotqa, 47761, 2017arXivtriviaqa} or the public internet \cite{bajaj2018msmarcohumangenerated}.  We discuss this further in \S \ref{sec:rag-benchmarking-related}.

Additionally, there is recent interest in privacy-preserving RAG \cite{zeng2024mitigatingprivacyissuesretrievalaugmented, zeng2024goodbadexploringprivacy} where it is important for a model to be able to assist and access knowledge from private documents without exposing personally identifiable information.  Personalized RAG over private documents \cite{10.1145/3626772.3657662, zerhoudi2024personaragenhancingretrievalaugmentedgeneration, 10.1007/978-981-97-0989-2_2} and federated information retrieval \cite{INR-010, 10.1145/3539618.3591926, 10.1145/3539618.3591926, 10.1145/3626772.3657853} require personal documents and segmented data stores.  Such explorations would benefit from a realistic corpus segmented into several private users for developing measured approaches to private and personalized RAG.

To serve these diverse tasks and increase coverage of benchmarking in private settings, we introduce the \benchmark benchmark based on the Enron emails corpus \cite{enron_email_dataset_2015}.  \benchmark contains 103,638 emails with 528,304 question-answer pairs, spanning 150 distinct user inboxes.  By designing a rigorous question generation pipeline grounded in specific evaluations we ensure a collection of high quality and diverse questions.  Our QA dataset is unmatched in size for openly available document retrieval over private documents and is large enough to enable finetuning, optimization, and evaluation over this setting.  Figure \ref{fig:teaser} showcases the \benchmark benchmark and some evaluation settings we test.

To showcase the utility of \benchmark we perform two case studies.  First, we showcase how benchmarking RAG pipelines on EnronQA has higher headroom for improving retriever quality (\ref{sec:}). We find that without retrieval, RAG pipelines score below 5\% on EnronQA, unlike other popular RAG benchmarks where, using parametric LLM knowledge, it's possible to score above 60\% without any retrieval at all.  Next we train LoRA adapters to memorize factual knowledge in our large dataset. Our LoRA adapters for memorization reveal that training LLMs to memorize private factual knowledge can perform on par with storing all facts in context, however retrieving the specific relevant information still outperforms both.

Overall, our contributions are as follows:

\begin{enumerate}
    \item The \benchmark benchmark, a collection of over 100,000 private emails and 500,000 questions, segmented into 150 distinct user inboxes (\S \ref{sec:dataset}).  
    \item We showcase the quality and utility of our benchmark and compare it with other popular RAG benchmarks (\S \ref{subsec:calibration}).  We benchmark popular RAG pipelines on \benchmark as a baseline for future work (\S \ref{sec:benchmarking}).
    \item We motivate memorizing private knowledge and showcase a LoRA-based method for memorizing factual knowledge, which performs competitively to putting a knowledge base in context (\S \ref{sec:memorization}).  Retrieving the most relevant information outperforms both, however we discuss future improvements that motivates further exploration into memorization adapters.
\end{enumerate}

\section{Related Work}

We organize our discussion of related work to span our core contributions: benchmarking RAG, and factual memorization in LLMs.

\subsection{Retrieval Augmented Generation Benchmarking}
\label{sec:rag-benchmarking-related}

\begin{table*}[]
\resizebox{\textwidth}{!}{%
\begin{tabular}{@{}lccll@{}}
\toprule
\textbf{Benchmark} & \textbf{Documents} & \textbf{QA Pairs} & \textbf{Domain} & \textbf{Source} \\ \midrule
ConcurrentQA \cite{arora2023reasoning} & 5.2M + 47k & 18.4k & General and Private Knowledge & Wikipedia + Emails \\
EManual \cite{nandy-etal-2021-question-answering} & 308k & 3.3k & Customer Support & TV Manual \\
HAGRID \cite{hagrid} & 32.8M & 2.6k & General Knowledge & Wikipedia \\
HotPotQA \cite{yang-etal-2018-hotpotqa} & 5.2M & 112.8k & General Knowledge & Wikipedia \\
INSCIT \cite{wu-etal-2023-inscit} & 6.6M & 4.7k & General Knowledge & Wikipedia \\
MS Marco \cite{bajaj2018msmarcohumangenerated} & 3.6M & 1.01M & General Knowledge & Web Pages \\
Natural Questions \cite{47761} & 5.9M & 323k & General Knowledge & Wikipedia \\
PubMedQA \cite{jin-etal-2019-pubmedqa} & 211.3k & 273.5k & Academic Research & Research Abstracts \\
QReCC \cite{qrecc} & 10M & 81k & General Knowledge & Web pages \\
SearchQA \cite{dunn2017searchqanewqadataset} & 6.9M & 140.5k & General Knowledge & Google Search \\
TechQA \cite{castelli2019techqadataset} & 802k & 1.4k & Customer Support & Tech Forums \\
TopiOCQA \cite{adlakha2022topiocqa} & 5.9M & 50k & General Knowledge & Wikipedia \\
TriviaQA \cite{joshi-etal-2017-triviaqa} & 662.7k & 96k & General Knowledge & Wikipedia \\ \midrule
\benchmark (Ours) & 103.6k & 528.3k & Private Knowledge & Emails \\ \bottomrule
\end{tabular}%
}
\caption{Comparison of document-based QA benchmarks. We limit to only resources with a corpus size above 50k documents for space, but provide a full comparison in Appendix \ref{sec:comparison-appendix}.  \benchmark covers a comparable corpus scale to many popular QA benchmarks while having vastly more QA pairs enabling training, optimization, and document memorization exploration.  Additionally, \benchmark spans the under explored private document domain using emails.}
\label{tab:comparison-small}
\end{table*}

We provide a brief list of popular RAG QA Benchmarks in Table \ref{tab:comparison-small} and a more comprehensive list in Table \ref{tab:comparison}.  Several common RAG benchmarks draw documents from Wikipedia \cite{joshi-etal-2017-triviaqa, yang-etal-2018-hotpotqa, 47761, adlakha2022topiocqa, hagrid, wu-etal-2023-inscit}.  This unfortunately makes the benchmarks less suitable for benchmarking RAG pipelines using modern LLMs which have memorized a lot of the contents of Wikipedia and general knowledge \cite{petroni-etal-2019-language}.

The most related work to ours is ConcurrentQA \cite{arora2023reasoning}, which creates a benchmark that relies on multihop reasoning over both the Enron emails corpus as well as Wikipedia.  We consider this an excellent resource in conjunction with ours, however it is worth noting that these benchmarks solve distinct problems.  For one, the ConcurrentQA benchmark limits to just one inbox, while \benchmark spans 150 distinct users, enabling the study of personalized RAG within the larger benchmark.  Second, ConcurrentQA has 18.4k QA pairs rather than the 528.3k in \benchmark, which makes our benchmark more suited for explorations involving fine-tuning factual knowledge, and continued pretraining.  \benchmark will enable the exploration of these emerging trends in information retrieval.  Finally, ConcurrentQA is comprised of multi-hop queries, which is very useful for benchmarking sophisticated pipelines.  We instead design \benchmark to be single hop to focus on the interesting cases of personalization and memorization, however in section~\ref{sec:corpus-filtering} we discuss how we make \benchmark fully compatible with ConcurrentQA so users can benchmark both single and multi-hop retrieval using our benchmark.


\subsection{Factual Memorization}

In our case study we explore factual memorization as an alternative to traditional retrievers for recalling facts. Factual memorization in LLMs is an exciting and relatively new direction.  Much of the work in LLM memorization comes from work studying unlearning \cite{maini2024tofu, liu2024rethinkingmachineunlearninglarge} in LLMs or understanding memorization from an interoperability lens \cite{huang2024demystifyingverbatimmemorizationlarge}. Some recent works have looked towards strategies for encouraging memorization in LLMs.  One approach is augmenting LLMs with external memory parameters \cite{collier-beel-2019-memory, graves2014neuralturingmachines}.  Other work encourages fine-tuning of LLMs for factual memorization \cite{lyu-etal-2024-knowtuning, roberts-etal-2020-much, tian2023finetuning}.

Perhaps the most relevant and exciting connection specifically to information retrieval is continued pretraining \cite{ke2023continual, gupta2023continualpretraininglargelanguage} where to adapt LLMs to new knowledge and domains, it is possible to just continue the pretraining process on more documents from the new source.  The most promising connection to our setting is "Synthetic Continued Pretraining" \cite{yang2025synthetic} wherein entities are extracted from documents, and connections are drawn between those entities. Then, the LLM is continuously pre-trained on these connections.  In doing so, the authors find that this encourages memorization of the documents, and ultimately, when tested with RAG, the performance benefits compound.  In this work, we release a benchmark of private and unmemorized documents to benchmark RAG performance.  Such a resource will be a rich test bed, providing a realistic QA task for continued pretraining methods.

\section{\benchmark Dataset Construction}
\label{sec:dataset}
We construct the \benchmark benchmark using the Enron emails corpus \cite{klimt2004enron}.  The Federal Energy Regulatory Commission released the original corpus during the Western Energy Markets investigation in 2003.  The original dataset contained over 600,000 messages and 158 distinct users (inboxes).  We use the 2015 release of the corpus, which has been cleaned and had emails removed at the request of Enron participants \cite{enron_email_dataset_2015}.  This version of the corpus contains 517,401 emails across 150 users.  To convert the raw emails into a high-quality RAG benchmark, we devise a 3-stage pipeline: Filtering (\S \ref{sec:corpus-filtering}), QA Generation (\S \ref{sec:qa-generation}), Post Processing (\S \ref{sec:post-processing}).

\subsection{Corpus Filtering}
\label{sec:corpus-filtering}
For our data filters, we take inspiration from popular pretraining data filtering pipelines \cite{dolma, together2023redpajama, pile, rae2022scalinglanguagemodelsmethods}.  Data filtering from raw emails draws many parallels to cleaning unstructured web data.  Table \ref{tab:corpus-filtering} outlines each step of the filtration process alongside the number of emails removed and example subjects of emails removed.

\begin{table*}[]
\resizebox{\textwidth}{!}{%
\begin{tabular}{@{}ccccl@{}}
\toprule
\textbf{Step} & \textbf{Operation} & \textbf{Documents} & \textbf{Change} & \textbf{Example Subjects of Affected Emails} \\ \midrule
0 & (Original) & 517,401 & -- & -- \\ \midrule
1 & Minhash Dedup & 228,098 & -289,303 & "daily charts and matrices as hot links 5/15", "Enron Mentions - 05/12/01 - 05/13/01" \\
2 & Subset Dedup & 192,759 & -35,339 & "FW: FW: Research Library", "Re: FW: Fedex", "Pre-Party??", "Re: Reference" \\
3 & Quality Filter & 103,513 & -89,246 & "RE: 606925 Finney Attasseril SRF (mar)", "Copyright Notice for EnronOnline" \\
4 & Language ID & 103,164 & -349 & "Praca dyplomowa", "Saludos, Feicitaciones y Follow up." \\
5 & NSFW Filter & 103,015 & -149 & "RE: bs", "dO yOU wANT tO pLAY wITH mE?", "Re: Richard Sucks!!!!!!" \\
6 & Toxicity Filter & 102,603 & -412 & "Re: LSU/Florida", "FW: Isn't it Funny", "I AM FEELING NEGLECTED" \\
7 & ConcurrentQA & 103,638 & +1,035 & "Internet Daily for November 16, 2001", "California Energy Crisis Timeline" \\ \bottomrule
\end{tabular}%
}
\caption{Enron Email Corpus throughout several steps of corpus filtering along with samples of subjects from emails added/removed at each step.}
\label{tab:corpus-filtering}
\end{table*}

\paragraph{Deduplication} Web data extraction pipelines use minhash deduplication \cite{leskovec2020mining} to reduce the number of identical documents and documents with minimal edits, such as software licenses.  In email inboxes, this can correspond with both email threads and email subscriptions.  For example, a subscription to a weather service might send the same email each day, with the forecast changed. We modify the text-dedup \cite{chenghao_mou_2023_8364980} library's minhash implementation to add a final step where the Jaccard similarity between matched documents is computed, and we run minhash deduplication with a Jaccard similarity threshold of 0.9, using 9 bands and 27 rows.  To handle thread deduplication, we also remove any emails that appear in their entirety within the content of another email which we call "subset" deduplication.

\paragraph{Gopher Quality Filters} \citep{rae2022scalinglanguagemodelsmethods} outline a few calculations to filter pretraining documents for quality.  They filter based on document word length, mean word length, number of lines ending in ellipsis, and the ratio of alphanumeric characters to symbols.  Tuning the cutoff for each of these rules for our email domain, we set the cutoffs to emails with between 50 and 1000 words, a length between 3 to 10, the ratio of alphanumeric characters to symbols greater than or equal to 0.65, and fewer than 10\% of lines ending in ellipsis.  This helps filter out excessively short emails and excessively long or low-quality emails, such as automated log files for software systems and automated financial reports.

\paragraph{Language Identification}  Some email threads between Enron employees occur in non-English languages.  Since we are creating an English benchmark using LLMs with limited multilingual training, we filter out documents classified as non-English or classified as English with below 80\% confidence with a fastText language identification model \cite{joulin2016bag}.

\paragraph{NSFW/Toxicity Filters}  Surprisingly, we identified some toxic and inappropriate content as the subject of some emails.  To preserve the participants' privacy and maintain professionalism in our \benchmark resource, we filter out these documents.  We use two fastText classifiers trained on jigsaw \cite{jigsaw-toxic-comment-classification-challenge} for the dolma project \cite{dolma}, one for detecting toxicity and one for detecting NSFW content.  We filter out emails that are not predicted as safe with greater than or equal to 90\% confidence.

\paragraph{ConcurrentQA} To make our resource compatible with the related ConcurrentQA \cite{arora2023reasoning} benchmark, we map all of the documents in the ConcurrentQA Enron corpus back to emails in our corpus.  If these rules filtered out any, we add them back in.

\subsection{QA Generation Pipeline}
\label{sec:qa-generation}
\begin{figure*}[h]
\begin{center}
\includegraphics[width=\textwidth]{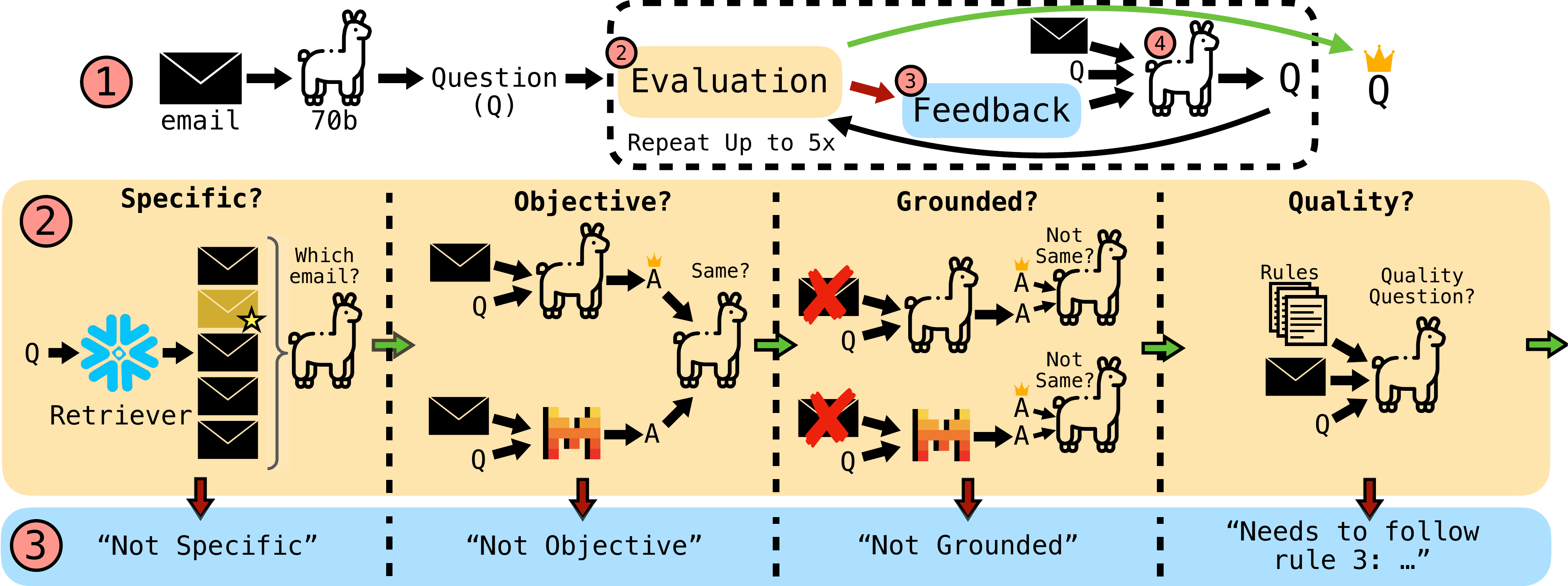}
\end{center}
\caption{Our multistage compound LLM system for QA Generation on the Enron emails corpus.  Our pipeline consists of 4 stages labeled in the diagram above: (1) Initial generation, (2) Evaluation, (3) Feedback generation, and (4) Refinement.  Producing a single high-quality question takes 10-50 distinct LLM calls, and the system is optimized end-to-end.  Our pipeline asserts that questions are specific, objective, grounded, and high-quality (correlating with human judgment).  All Llama icons correspond to Llama3.1 70b Instruct \cite{dubey2024llama3herdmodels}, the Mistral icon represents the Mixtral-7B-Instruct model \cite{jiang2024mixtralexperts}, and our retriever is a bi-encoder using Snowflake's artic-embed-m-v1.5 \cite{merrick2024embeddingclusteringdataimprove}.}
\label{fig:qa-generation}
\end{figure*}

We must generate high-quality questions to convert the cleaned corpus into a RAG benchmark.  We devise a multi-stage compound LLM system implemented and optimized end-to-end in DSPy \cite{khattab2023dspy}.  The generation of a single question comprises between 10 and 50 distinct LLM calls each designed to serve a single modular purpose.  Our pipeline is described visually in Figure \ref{fig:qa-generation}, and can be conceptualized as 4 main stages: (1) Initial generation, (2) Evaluation, (3) Feedback generation, (4) Refinement.

\paragraph{Initial Generation} We generate an initial question given a document and a set of prior questions for that document (so that the LLM does not generate repeats).  Our question generator model is llama3.1-70b \cite{dubey2024llama3herdmodels} with a prompt optimized by DSPy using the MIPROv2 optimizer \cite{opsahlong2024optimizinginstructionsdemonstrationsmultistage}.  Our optimization objective was to reduce the number of refinement steps from our early stopping pipeline.  The pipeline repeats if any of the \textit{Evaluation} steps described below fail.  Our optimized prompts brought the pipeline from an average of 1.94 repetitions down to 1.64 repetitions.  We include our initial and optimized prompt in appendix \ref{sec:initial-qa-generation-prompt}.

\paragraph{Evaluation} Given a question, we want to assess whether the question is high quality.  To do this, we introduce four evaluation criteria and concrete measures: Specificity, Objectivity, Groundedness, and Quality.

\begin{itemize}
    \item[] \textbf{Specificity.} We designate a question as "specific" if given ten similar emails (including the true email the question is about) an LLM can pick out which email would answer the question.  We mine hard negative examples by retrieving the top 10 relevant documents from our corpus given the question.  We use a biencoder built on Snowflake arctic-embed-m-v1.5 \cite{merrick2024embeddingclusteringdataimprove} to retrieve the top 10 most similar documents.  We use Llama3.1 70b as our selector LLM.  The full prompt is provided in Appendix \ref{sec:email-selection-prompt}.
    \item[] \textbf{Objectivity.} We determine a question to be "objective" if two models from different families answer the same question with the same answer, given the email as context.  Here we use Llama3.1 70b Instruct \cite{dubey2024llama3herdmodels} and Mixtral 8x7B Instruct \cite{jiang2024mixtralexperts}.  We use an LLM as a judge to determine if the answers match.  Our Llama3.1 70B Instruct LLM judge achieves a 0.98 F1-score with human evaluation on a small study of 200 generations.  We include all QA prompts in Appendix \ref{sec:question-answering-prompts} and details of our LLM as a judge evaluation in Appendix \ref{sec:llm-as-a-judge-prompt}.  Given the question is deemed objective, we save the Llama3.1 70B Instruct output as the "gold answer."
    \item[] \textbf{Groundedness.} We determine a question to be "grounded" if neither the llama nor mixtral model can answer the same as the gold answer given no email as context.  This both tests that the answers to our questions are not memorized and that the questions are not easily guessable.  Again, we use the LLM as a judge, grounded in the email, to determine if the answers match the "gold answer" obtained from the previous evaluation.  If neither ungrounded answer matches the gold answer, we deem the question "grounded."  All QA prompts are included in Appendix \ref{sec:question-answering-prompts}.
    \item[] \textbf{Quality.} Our last evaluation step is measuring the "quality" of the question by an LLM judge aligned with human judgments.  We generated 20 questions using the pipeline with only the specificity, objectivity, and groundedness stages. We had two authors label them as "high," "medium," or "low" quality based on a rubric assessing specificity, objectivity, and groundedness.  The authors had a 0.5 Spearman correlation of their annotations, and a third author adjudicated the disagreements.  That same author also independently labeled 21 more questions as "high", "medium", or "low".  Using the 21 singly-labeled questions as a development set and the 20 group-labeled questions as a test set, we devised a list of rules for Llama-3.1 70B Instruct to use to determine if a question was "high" or "low" quality (wrapping the "medium" label into low quality).  Our ruleset enabled the judge to achieve 85.7\% accuracy on the development set, and running it once on the test set yielded 85\% accuracy.  The final stage of our evaluation pipeline uses a Llama3.1 70B Instruct model augmented with a ruleset to determine if the question is "high quality."  We include our ruleset in Appendix \ref{sec:quality-evaluation-prompts}.
\end{itemize}

\paragraph{Feedback Generation} Based on the latest stage that the question made it to in the evaluation phase, we produce feedback to add to the context of the refinement step.  

\begin{itemize}
    \item If the question is not specific, we handle this in a special case described in the "Refinement" step.
    \item If the question is not objective, we provide the feedback: "Question is not objective.  Different annotators answer the same question differently given the same email as context.  Could benefit from more clarity."
    \item If the question is not grounded, we provide the feedback: "Question is not grounded.  It is too easy to guess the answer to this question without having read the email."
    \item If the question fails the quality check, we use the chain-of-thought reasoning of the LLM as the feedback for why the question is not of high quality.  This typically cites which rule the email fails, briefly explaining why.
\end{itemize}      

\paragraph{Refinement} If our question succeeds at all evaluation stages, it is considered a good question and added to our question bank.  Otherwise, we need to refine it.  We have two refinement steps.  If the question is not specific, we show the LLM the ten retrieved emails from the specificity check and ask it to rewrite the question to be more specific to only the gold email.  If the question fails the other steps, we use our generated feedback to ask the LLM to rewrite the question and address the feedback.  Both the specificity and general feedback rewrite prompts are optimized using DSPy.  We include the initial and optimized specificity refinement prompt in Appendix \ref{sec:refinement-specificity-prompt} and the initial and optimized feedback prompt in Appendix \ref{sec:refinement-feedback-prompt}.

\subsection{Additional Data Processing}
\label{sec:post-processing}

\begin{figure}
  \begin{center}
    \includegraphics[width=0.48\textwidth]{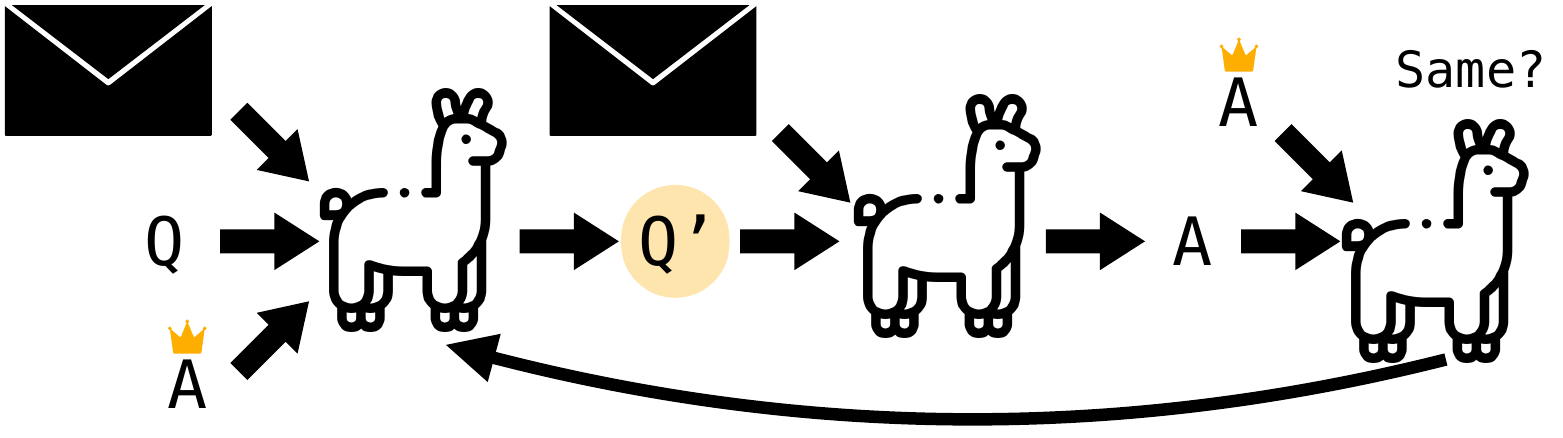}
  \end{center}
  \caption{Question Rewrite Pipeline.  First, we ask Llama 3.1 70B Instruct to rewrite the question, and then we ask it to answer this new question.  Finally, we use Llama 3.1 70B Instruct to check that the answers match.\label{fig:qa-rephrase}}
\end{figure}

We provide a rewritten version of our questions to make our dataset more practical for various downstream tasks, such as our memorization case study (\S \ref{sec:memorization}).  In the case of training LLMs to memorize specific information, this enables you to train and test different questions while retaining the informational content.  In Figure \ref{fig:qa-rephrase}, we showcase our pipeline for rephrasing questions.  We use Llama3.1-70B-Instruct to rewrite the question, answer the rewritten question, and finally judge if the answer is the same.  If the answers don't match, we try again up to 15 times before discarding the question.  We discard 265/528,569 questions in this process.

Alongside the core components of our dataset: the questions, gold answers, emails, and rephrased questions, we also release miscellaneous artifacts produced in creating the core dataset.  These artifacts include the verified answers of the Mixtral-8x7B-Instruct model from the evaluation step, as well as the chain of thought reasoning for both Llama3.1-70b-Instruct and Mixtral-8x7B-Instruct as they answer each question in the \benchmark benchmark conditioned on the oracle document.

\section{Dataset Quality}

We discuss here some of the properties of the \benchmark benchmark and what makes it a high quality and valuable resource to the community.

\subsection{Dataset Statistics}

We report summary statistics for the \benchmark benchmark in Table~\ref{tab:stats}.  Notably the benchmark contains over 333k training questions, and the median number of questions about a single user's emails is over 1k.  The \benchmark benchmark is suitably large for fine-tuning on questions, continuously pretraining on documents, and benchmarking RAG pipelines.

\begin{table}[htbp]
  \centering
  \small
  \begin{tabular}{lccc}
    \toprule
    \textbf{Metric} & \textbf{Train} & \textbf{Dev} & \textbf{Test} \\
    \midrule
    \multicolumn{4}{l}{\textit{Email Stats (consistent across splits)}} \\
    \midrule
    Average emails per user   &   & 491.81  &  \\
    Median emails per user    &    & 240.5   &   \\
    Mean email length (chars) & & 2,269.69 & \\
    Median email length (chars)& & 1,664.0  &  \\
    \midrule
    \multicolumn{4}{l}{\textit{Question Stats}} \\
    \midrule
    Mean questions per user   & 2,223.15 & 703.43  & 595.44  \\
    Median questions per user & 1,063.0  & 334.5   & 278.0   \\
    Mean questions per email  & 4.52    & 1.43    & 1.21    \\
    Median questions per email& 3.0     & 1.0     & 1.0     \\
    Total questions count     & 333,473  & 105,515  & 89,316   \\
    \bottomrule
  \end{tabular}
  \caption{Summary statistics \benchmark benchmark.  The benchmark contains suitably large amounts of documents and questions for continued pertaining and RAG benchmarking.}
  \label{tab:stats}
\end{table}

\subsection{Calibration}
\label{subsec:calibration}

One downside to using common RAG benchmarks, which pull documents from Wikipedia, is the lack of calibration between benchmark scores and retrieval quality.  A comparative advantage to \benchmark is that, for the most part, the parametric knowledge encoded in the LLMs will not memorize the Enron emails.  Thus, you can expect gains from a better retriever to match gains in accuracy on the benchmark.  To test this hypothesis, we choose two standard RAG benchmarks \textbf{NaturalQuestions} \cite{47761} and \textbf{TriviaQA} \cite{2017arXivtriviaqa}.  NaturalQuestions comprises 323,000 queries to Google, with answers spanning 5.9M Wikipedia documents.  TriviaQA contains 95K question-answer pairs authored by trivia enthusiasts and over 600 thousand articles.  We specifically use the KILT \cite{petroni-etal-2021-kilt} versions of the datasets.

\begin{figure*}[h!]
\begin{center}
\includegraphics[width=\textwidth]{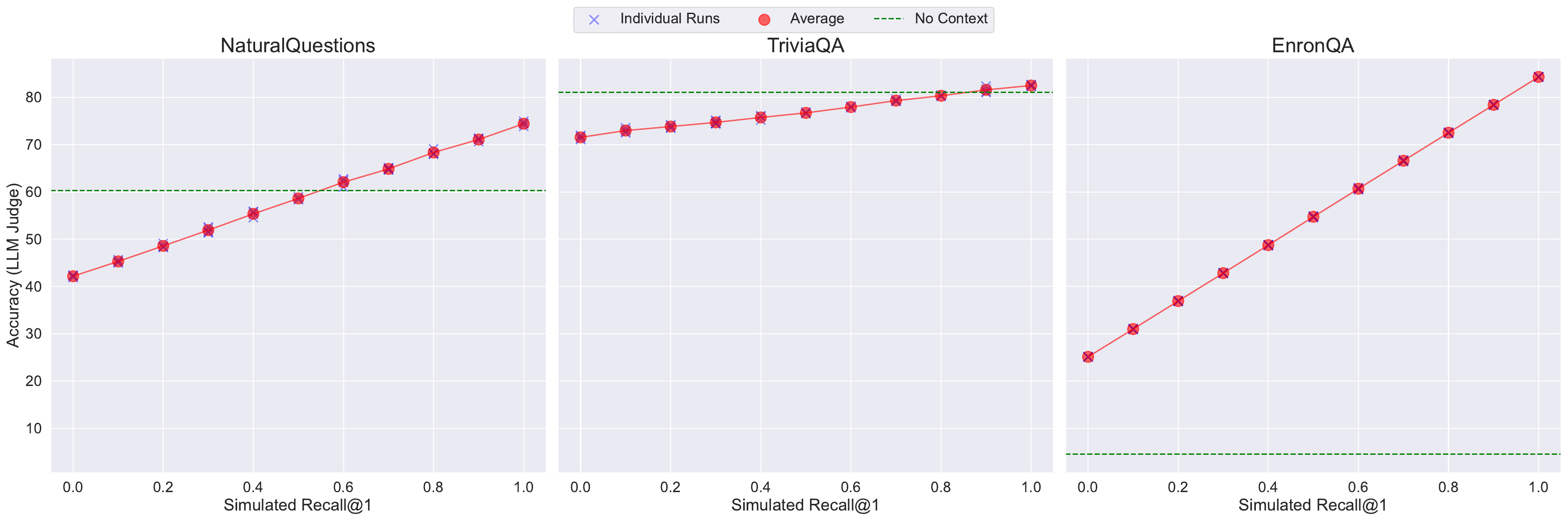}
\end{center}
\caption{Calibration experiment results.  Although all benchmarks scale roughly linearly with more accurate context, \benchmark is the only benchmark where adding context always outperforms the no-context baseline.  For TriviaQA, it takes Recall@1 of nearly 0.85 to surpass the performance of the no-context baseline.}
\label{fig:calibration}
\end{figure*}

\paragraph{Experimental Setting} We filter to 10,000 training / 500 validation examples for NaturalQuestions and TriviaQA.  And 1,000 training / 500 validation examples for EnronQA.  We use the full validation sets of NaturalQuestions and TriviaQA as the test set.  For EnronQA, we use the actual test set.  We optimize two DSPy programs for each setting using the MIPROv2 optimized \cite{opsahlong2024optimizinginstructionsdemonstrationsmultistage}.  The first program takes no context and has to answer the question directly.  The second program takes the gold document as context and answers the question.  We optimize with Llama-3.1-8B-Instruct as our task model and Llama-3.1-70B-Instruct as our prompt model with 10 candidate programs.  When running the experiment, we use Llama-3.1-70B-Instruct.  We run the no-context case; then, we simulate Recall@1 between 0.0 and 1.0 by randomly including the correct document as context or a randomly sampled document instead.  We tested with five random seeds and averaged the results.  Scores are produced using Llama-3.1-70B-Instruct as a judge for answer accuracy compared to the gold answer.

\begin{table*}[h!]
\resizebox{\textwidth}{!}{%
\begin{tabular}{@{}rcccccccccc@{}}
\toprule
 &  & \multicolumn{3}{c}{No Query Rewrite (acc)} & \multicolumn{3}{c}{Query Rewrite (acc)} & \multicolumn{3}{c}{Query Rewrite (R@5)} \\ \midrule
\multicolumn{1}{r|}{} & \multicolumn{1}{c|}{R@5} & Llama8B & Llama70b & \multicolumn{1}{c|}{GPT4o} & Llama8B & Llama70b & \multicolumn{1}{c|}{GPT4o} & Llama8B & Llama70b & GPT4o \\ \midrule
\multicolumn{1}{r|}{BM25} & \multicolumn{1}{c|}{87.5} & 72.1 & 77.1 & \multicolumn{1}{c|}{81.2} & 66.1 & 80.0 & \multicolumn{1}{c|}{81.1} & 82.5 & 86.6 & 88.5 \\
\multicolumn{1}{r|}{ColBERT} & \multicolumn{1}{c|}{54.1} & 65.8 & 74.6 & \multicolumn{1}{c|}{74.9} & 56.3 & 69.1 & \multicolumn{1}{c|}{73.3} & 54.1 & 59.4 & 59.3 \\ \bottomrule
\end{tabular}%
}
\caption{Benchmarking several retrieval methods and LLMs on the \benchmark benchmark both with and without query rewriting.  Surprisingly, simple retrieval baselines (BM25) work well on our benchmark.  This is likely due to some lexical overlap between the queries and proper nouns in the emails, such as names and events.}
\label{tab:benchmarking}
\end{table*}

\paragraph{Results} We include results of the experiment in Figure~\ref{fig:calibration}.  We find that \benchmark is the only benchmark where adding context is always better than the no-context baseline.  For NaturalQuestions it takes a retriever with a Recall@1 above 0.5 in order to outperform the no context baseline.  Likewise on TriviaQA the problem is even worse.  Just asking Llama-3.1-70B-Instruct the question directly without context outperforms all retrieval-based systems with Recall@1 less than 0.85!  This means that any accuracy changes on RAG pipelines benchmarking on TriviaQA with Recall@1 less than 0.85 may have more to do with the memorized knowledge of the LLM rather than retrieval quality.

In contrast, with \benchmark all improvements with more accurate context lead directly to higher accuracy on the benchmark. Additionally, \benchmark showcases the highest improvement in accuracy for every point of recall gain.  Nearly a 0.6\% gain in accuracy for every 1\% increase in recall.  This is because the knowledge in \benchmark has not been memorized by large foundation models, which is the problem that trivializes NaturalQuestions and TriviaQA.

\section{Benchmarking RAG Pipelines}
\label{sec:benchmarking}

To offer some baseline performance numbers and show off the utility of \benchmark for RAG benchmarking, we test a sweep of two popular retrievers, three popular LLMs, and two common RAG pipeline architectures.

\subsection{Experimental Setting}

\paragraph{Retrievers} We test \textbf{BM25} using the PySerini implementation \cite{Lin_etal_SIGIR2021_Pyserini} and \textbf{ColBERTv2} \cite{santhanam-etal-2022-colbertv2} over the full set of 103,638 emails.  We retrieve five documents simultaneously for a single call to the retriever.  For each retriever, we also report Recall@5.

\paragraph{Large Language Models} We test with \textbf{Llama-3.1-8B-Instruct}, \textbf{Llama-3.1-70B-Instruct} \cite{dubey2024llama3herdmodels}, and \textbf{GPT4o} to test models of different scales and families.

\paragraph{RAG Architectures} We test two RAG settings; first, we test \textbf{No Query Rewrite}, where we search the query directly using the question directly from \benchmark.  We then provide the top 5 results from the retriever and pass the retrieved documents and questions to the LLM to be answered.  We additionally test the \textbf{Query Rewrite} setting where we first have the LLM rewrite the question into a search query.  Then, we retrieve five emails.  Finally, given the five emails and the question, we have the LLM produce the answer.  Prior works have found query rewriting with LLMs to help with adapting to the specifics of a particular retriever \cite{ma-etal-2023-query} so we test this on our benchmark.  For both settings and with all models and retrievers, we optimize the prompts and few-shot demonstrations using DSPy MIPROv2 \cite{opsahlong2024optimizinginstructionsdemonstrationsmultistage} with 10 candidates and 20 trials.

\begin{table*}[h!]
\resizebox{\textwidth}{!}{%
\begin{tabular}{@{}rcccccccclll@{}}
\toprule
\multicolumn{1}{c}{\# Facts $\downarrow$} & Long Context & RAG & \multicolumn{9}{c}{Memorization (LoRA)} \\ \midrule
\multicolumn{1}{c|}{Rank $\rightarrow$} & - & \multicolumn{1}{c|}{-} & 8 & 16 & 32 & 64 & 128 & 256 & 512 & 1024 & 2048 \\ \midrule
\multicolumn{1}{r|}{10} & 0.80 & \multicolumn{1}{c|}{1.00} & 0.80 & 0.80 & 0.80 & 0.80 & 0.80 & 0.80 & 0.90 & 0.80 & 0.80 \\
\multicolumn{1}{r|}{100} & 0.91 & \multicolumn{1}{c|}{0.95} & 0.76 & 0.76 & 0.80 & 0.84 & 0.83 & 0.85 & 0.87 & 0.88 & 0.78 \\
\multicolumn{1}{r|}{500} & 0.83 & \multicolumn{1}{c|}{0.91} & 0.75 & 0.73 & 0.79 & 0.81 & 0.78 & 0.82 & 0.79 & 0.80 & 0.80 \\
\multicolumn{1}{r|}{1000} & 0.79 & \multicolumn{1}{c|}{0.89} & 0.72 & 0.71 & 0.79 & 0.79 & 0.79 & 0.78 & 0.76 & 0.78 & 0.73 \\
\multicolumn{1}{r|}{5000} & NA & \multicolumn{1}{c|}{0.92} & 0.53 & 0.55 & 0.69 & 0.75 & 0.74 & 0.69 & 0.78 & 0.03 & \multicolumn{1}{c}{---} \\
\multicolumn{1}{r|}{10000} & NA & \multicolumn{1}{c|}{0.92} & 0.53 & 0.61 & 0.69 & 0.74 & 0.77 & 0.69 & 0.78 & 0.79 & \multicolumn{1}{c}{---} \\
\multicolumn{1}{r|}{20000} & NA & \multicolumn{1}{l|}{0.93} & 0.53 & 0.62 & 0.69 & 0.74 & 0.75 & 0.08 & 0.00 & 0.03 & \multicolumn{1}{c}{---} \\ \bottomrule
\end{tabular}%
}
\caption{Factual memorization on subset of \benchmark benchmark.  While currently, RAG is the best-performing method of recalling factual information, training LoRA adapters for memorization can match the performance of putting all the facts in context, suggesting this is a promising direction for future development.}
\label{tab:memorization}
\end{table*}

\subsection{Results}

We present results in Table~\ref{tab:benchmarking}.  We find surprisingly high accuracy from the BM25 retriever, boasting a Recall@5 of 87.5 without any additional query rewrite steps.  This is likely due to high lexical overlap between some of the queries and the email contents.  Because our pipeline for question generation was optimized to be specific enough to pick one email out of a batch of ten, the queries had to name particular entities within the emails.  This is reflected in the high BM25 accuracy.  We find that unsurprisingly larger models get better at our benchmark with performance scaling from 8b to 70b to GPT4o.  We also find that query rewriting was not particularly helpful for this benchmark, especially for BM25.  The highest performing setting was GPTo, both with and without query rewrites using BM25, which achieved an accuracy of 81.2\% on \benchmark.



\section{Case Study: Memorized Knowledge}
\label{sec:memorization}

With a growing body of literature on continued pre-training \cite{yang2025synthetic}, we note that an interesting use case of our benchmark is a large-scale and realistic test bed for continued pre-training memorization.  Since our benchmark contains private knowledge that LLMs have not been heavily pretrained on, alongside over 500k question and answer pairs, there is plenty of data to benchmark and even fine-tune models on to test parametric knowledge memorization, and to benchmark this against RAG.

To this end, we provide initial results in this direction, hoping that this resource will be useful to future researchers exploring continued pretraining and memorization with LLMs.

\subsection{Experimental Setting} We want to explore the memorization/retreival of between 10 and 20,000 facts about documents by three mechanisms: \textbf{Long Context}, \textbf{RAG}, and \textbf{Memorization}.  For this setting, we simplify the problem by looking at question-and-answer pairs directly rather than the documents, though we hope future work can also explore training on the documents.  We use the rephrased question and answer pair as the context set, and tested on the true question and answer pair.  For Long Context, we put all the QA-pairs (facts) we were trying to memorize in the context alongside their answers in the context of Llama-3.1-8B-Instruct.  We could test as high as 1,000 QA-pairs until the context length was full.  For RAG, we build an index over all the QA-pairs and retrieve the top 100 (selected because it was the best for Long Context) most relevant question-answer pairs to the context of Llama-3.1-8B-Instruct.  We use ColBERTv2 \cite{santhanam-etal-2022-colbertv2} as our retreiver.  Finally, for Memorization, we train a LoRA adapter using the setup from the "Task of Fictitious Unlearning" paper, which tests unlearning on LoRA adapters \cite{maini2024tofu}.  We train LoRA adapters of rank $\{8,\,16,\,32,\,64,\,128,\,256,\,512,\,1024,\,2048\}$ on all the facts for 10 epochs with rate~$1\times10^{-4}$.  We set alpha to four times the rank and used a dropout of 0.05.  We test ablations with which layer to adapt, and find that doing all linear layers works the best.  All settings are evaluated with LLama-3.1-70B-Instruct as a judge.

\subsection{Results} We present the results of this experiment in Table~\ref{tab:memorization}.  Interestingly, LoRA memorization can match long-context performance at almost all scales and continue beyond the 1000 QA-pair cap that blocks long-context from scaling.  In fact, for many of the LoRA adapters, the performance only starts to degrade around 20,000 facts memorized, showing a surprising capacity packed into just the LoRA parameters.  At all scales, RAG outperforms memorization and long-context. This is likely due to the simplicity of the task (retrieving a rephrased QA pair) as well as the strength of current RAG systems.  Memorization is a relatively understudied phenomenon (mostly explored with LLMs to try to \textit{prevent} memorization), so, unsurprisingly, this does not yet outperform RAG.  In the future, with the continued development of pretraining and memorization methods, it is possible that memorization through LoRA adapters could match or exceed RAG performance.

\section{Discussion}

Here, we discuss some lessons learned and valuable insights for researchers working on similar problems.

\textbf{LLM self-verifying and optimizing pipelines can be powerful synthetic data tools.}  Our \benchmark benchmark is comprised of entirely synthetically generated question-and-answer pairs.  Past processes of generating such QA resources would require a massive human undertaking or need to be crowdsourced from platforms where people naturally ask questions, such as Google \cite{47761} or Bing \cite{bajaj2018msmarcohumangenerated}.  Instead with the growing capabilities of LLMs we were instead able to specify the requirements of our questions and answers into verifiable unit tests. The questions needed to be "Specific," "Objective," "Grounded," and "High Quality."  By writing each of these checks as unit tests and optimizing our system end-to-end to pass these unit tests, we were able to synthetically generate a large scale dataset while maintaining quality.  Questions only made our final benchmark if they passed through all four of the unit tests successfully.  We recognize this as an extensible pattern: (1) write specifications into unit tests, (2) optimize pipeline (fine-tuning, prompts, etc.), (3) filter synthetic generations based on unit tests.  This will be a way to scale up data collection efforts in the future, which will be heavily reliant on the design of the unit tests themselves.

\textbf{Memorization through fine-tuning or continued pretraining are interesting future directions for retrieval.} The current SOTA for RAG is to retrieve and then pass this context to an LLM.  We showed, however, that LLMs are capable of memorizing large amounts of data.  For example, past RAG benchmarks like NaturalQuestions \cite{47761} and TriviaQA \cite{2017arXivtriviaqa} have all been consumed by the parametric knowledge of LLMs.  Right now this parametric knowledge is largely dictated by the composition of the internet which is the largest source of training data for these models.  In the future one could imagine doing continued pretraining on private documents or an additional fine-tuning step for memorization.  In section~\ref{sec:memorization}, we show some first steps towards this effort and find that LoRA adapters can match long-context at recalling factual knowledge in a simplified setting.  With more work on continued pretraining, we hope that \benchmark can serve as a resource for testing these sorts of methods and exploring the limits of LLM memorization in the future. 

\section{Conclusion}
We introduce \benchmark, a dataset of 103,638 emails with 528,304 question-answer pairs across 150 different user inboxes.  \benchmark enables better benchmarking of RAG pipelines over private data and allows for experimentation on the introduction of personalized retrieval settings over realistic data.  We showed that the \benchmark benchmark is better than other single-hop retrieval benchmarks for measuring the joint accuracy of retrievers and LLMs.  We benchmark existing RAG pipelines over a sweep of retrievers, LLMs, and architectures on \benchmark.  Finally, we use \benchmark to explore the tradeoff in memorization and retrieval when reasoning over private documents.  We release this large resource publicly to the community for testing private and personalized retrieval and to enable further research in continued pretraining, which is a potential new frontier for information retrieval from the parametric weights of Large Language Models.

\subsection*{Ethics Statement}
The \benchmark benchmark draws from the Enron emails corpus \cite{klimt2004enron}, which was a release of corporate emails as a part of the Western Energy Markets investigation in 2003.  Not all Enron employees whose emails were released were guilty of any crimes, and even still, we wish to respect the wishes of all the humans behind the Enron emails regardless of involvement in the criminal activity.

We take two critical steps to support these goals in respecting the Enron employees behind the dataset.  First, we use the 2015 release of the dataset where several people were removed from the dataset upon request \cite{enron_email_dataset_2015}.  Second, we apply a filter to remove any NSFW or toxic content from the dataset (\S\ref{sec:corpus-filtering}), which can be particularly personal.

Beyond this, we are more than happy to support any requests for data removal from any affected parties.  The \benchmark dataset will be continuously maintained and updated should any such removal requests arise.  The Enron emails dataset has been used for about twenty years in academic research, and we hope to support the continued ethical use of this resource.


\bibliographystyle{ACM-Reference-Format}
\bibliography{custom}

\appendix

\section{Comparison with other QA and RAG benchmarks}
\label{sec:comparison-appendix}
Table \ref{tab:comparison} contains a comparison with other popular QA and RAG benchmarks.  \benchmark covers the under explored private knowledge domain using private emails.  It has a comparable or larger number of documents to other resources while covering vastly more questions.  Having multiple questions per document will facilitate training memorization of factual information in documents, and enables research finetuning and optimizing RAG pipelines rather than just serving as a diagnostic benchmark.

\begin{table*}[]
\resizebox{\textwidth}{!}{%
\begin{tabular}{@{}lccll@{}}
\toprule
Benchmark & Corpus Size & QA Pairs/Turns & Domain & Source \\ \midrule
ConcurrentQA \cite{arora2023reasoning} & 5.2M + 47k & 18.4k & General + Private Knowledge & Wikipedia + Emails \\
ConvFinQA \cite{chen-etal-2022-convfinqa} & 2k & 14k & Finance & Finance Reports \\
CoQA \cite{reddy-etal-2019-coqa} & 8.4k & 127k & General Knowledge & \begin{tabular}[c]{@{}l@{}}Literature, Academia, News,\\ Wikipedia, Reddit, Exams\end{tabular} \\
CovidQA \cite{moller-etal-2020-covid} & 147 & 2k & Academic Research & Research Papers \\
CUAD \cite{hendrycks2021cuadexpertannotatednlpdataset} & 510 & 13k & Legal & Legal Contracts \\
DelucionQA \cite{sadat-etal-2023-delucionqa} & 1 & 2k & Customer Support & Jeep Manual \\
Doc2Dial \cite{feng-etal-2020-doc2dial} & 458 & 25.7k & Government & Government Sites \\
DoQA \cite{campos2020doqaaccessingdomainspecific} & 2.4k & 10.9k & Cooking, Travel, Movies & Stack Exchange \\
EManual \cite{nandy-etal-2021-question-answering} & 308k & 3.3k & Customer Support & TV Manual \\
ExpertQA \cite{malaviya2024expertqaexpertcuratedquestionsattributed} & -- & 2.2k & Expert Knowledge & Google Search \\
FinQA \cite{chen-etal-2021-finqa} & 2.8k & 8.3k & Finance & Finance Reports \\
HAGRID \cite{hagrid} & 32.8M & 2.6k & General Knowledge & Wikipedia \\
HotPotQA \cite{yang-etal-2018-hotpotqa} & 5.2M & 112.8k & General Knowledge & Wikipedia \\
HybriDial \cite{nakamura-etal-2022-hybridialogue} & 2.9k & 22.5k & General Knowledge & Wikipedia \\
INSCIT \cite{wu-etal-2023-inscit} & 6.6M & 4.7k & General Knowledge & Wikipedia \\
MS Marco \cite{bajaj2018msmarcohumangenerated} & 3.6M & 1.01M & General Knowledge & Web Pages \\
NarrativeQA \cite{kocisky-etal-2018-narrativeqa} & 1.6k & 46.8k & Movie Scripts, Literature & \begin{tabular}[c]{@{}l@{}}Project Gutenberg\\ + IMSDB\end{tabular} \\
Natural Questions \cite{47761} & 5.9M & 323k & General Knowledge & Wikipedia \\
NewsQA \cite{trischler-etal-2017-newsqa} & 12.7k & 119.6k & News & CNN \\
PubMedQA \cite{jin-etal-2019-pubmedqa} & 211.3k & 273.5k & Academic Research & Research Abstracts \\
QReCC \cite{qrecc} & 10M & 81k & General Knowledge & Web pages \\
QuAC \cite{choi-etal-2018-quac} & 8.9k & 98.4k & General Knowledge & Wikipedia \\
SearchQA \cite{dunn2017searchqanewqadataset} & 6.9M & 140.5k & General Knowledge & Google Search \\
SQA \cite{iyyer-etal-2017-search} & 2.1k & 17.6k & General Knowledge & Wikipedia Tables \\
Squad 2.0 \cite{rajpurkar2018knowdontknowunanswerable} & 536 & 151k & General Knowledge & Wikipedia \\
TAT-QA \cite{zhu-etal-2021-tat} & 182 & 16.6k & Finance & Finance Reports \\
TechQA \cite{castelli2019techqadataset} & 802k & 1.4k & Customer Support & Tech Forums \\
TopiOCQA \cite{adlakha2022topiocqa} & 5.9M & 50k & General Knowledge & Wikipedia \\
TriviaQA \cite{joshi-etal-2017-triviaqa} & 662.7k & 96k & General Knowledge & Wikipedia \\
UDA \cite{hui2024udabenchmarksuiteretrieval} & 3k & 29.6k & \begin{tabular}[c]{@{}l@{}}Finance, Academia, \\ Knowledge Bases\end{tabular} & \begin{tabular}[c]{@{}l@{}}Finance, Research Papers,\\ Wikipedia\end{tabular} \\ \midrule
\benchmark (Ours) & 103.6k & 528.3k  & Private Knowledge & Emails \\ \bottomrule
\end{tabular}%
}
\caption{Comparison of document based QA benchmarks.  \benchmark covers a comparable or larger corpus scale to many popular QA benchmarks while having vastly more QA pairs enabling training, optimization, and document memorization exploration.  Additionally \benchmark spans the under explored private document domain using emails.}
\label{tab:comparison}
\end{table*}

\onecolumn
\section{Language Model Prompts}

\subsection{Initial QA Generation Prompt}
\label{sec:initial-qa-generation-prompt}
Here, we include both the unoptimized prompt for QA generation and the DSPy MIPROv2 \cite{opsahlong2024optimizinginstructionsdemonstrationsmultistage} optimized prompt, including a rewritten instruction and four bootstrapped few shot examples.  This prompt is to seed the question refinement process by creating an initial question based on the email and distinctive of the prior questions.  We optimize with a training set of thirty emails and 20 validation emails.  We run MIPROv2 for 20 iterations (batches) and generate 10 candidate instructions to search over.  We use Llama3.1 70b Instruct \cite{dubey2024llama3herdmodels} as our prompt generator model.
    

\begin{tcolorbox}[colback=blue!5!white, colframe=blue!75!black, title=Initial QA Generation Prompt (Unoptimized)]
Given a particular email and a set of prior questions I've already asked about this email, write a new question that tests the reader's understanding of the email.  The question should be specific to the factual contents of this email, and answerable with a single sentence, and should not be a repeat of any of the prior questions.\\
\\
---\\
\\
Follow the following format.\\
\\
Email: The email we want to test the reader's comprehension of\\
Prior Questions: The prior questions I've already asked about this email\\
Reasoning: Let's think step by step in order to \$\{produce the new question\}. We ...\\
New Question: The brand new question we want to ask about the email\\
\\
---\\
\\
Email: \{\textbf{email}\}\\
Prior Questions: \{\textbf{prior questions}\}\\
Reasoning: Let's think step by step in order to
\end{tcolorbox}

\begin{tcolorbox}[colback=blue!5!white, colframe=blue!75!black, title=Initial QA Generation Prompt (DSPy Optimized) (Part 1)]
Imagine you are a high-stakes business consultant, and you must generate a question that tests the reader's understanding of this email in order to provide a crucial bit of information needed to close a multi-million dollar deal. You should skew the question to be *entirely factual*, i.e. do not offer opinions, test the reader's own thoughts on matters, or request the emails conclusion. Do not suggest that you are confused. Additionally, the question cannot be answerable with a subjective judgment call. However, it is okay to use an interrogative that is compound, e.g. 'What is X, and what is Y?' Writing such a question quickly and to-the-point will mean receiving a substantial bonus. Given a particular email and a set of prior questions I've already asked about this email, write a new question to ensure success.\\
\\
---\\
\\
Follow the following format.\\
\\
Email: The email we want to test the reader's comprehension of\\
Prior Questions: The prior questions I've already asked about this email\\
Reasoning: Let's think step by step in order to \$\{produce the new\_question\}. We ...\\
New Question: The brand new question we want to ask about the email\\
\\
---\\
\\
Email: Subject: LUNCH\\Sender: buylow@houston.rr.com\\Recipients: ['dfarmer@enron.com']\\File: farmer-d/personal/292.\\=====================================\\TREEBEARD'S \\\\I am hooked. Name the time and place to  meet......................\\\\Tell Ami good luck for me. \\\\Who would have thunk OU would be back so quickly. Now we have  to contend \\with them, K-State, and the conference championship (if we survive).  I don't \\see the Big 12 being \#1 if we don't lighten up on each  other.\\\\You want a chuckle----- My daughter is a big NU fan and has  gone to a couple \\of Baylor games. Says it's not quite the same as Lincoln;  but the \\expectations are not high.\\\\Let me know on lunch and I'll be there.\\\\Ken\\=====================================\\
Prior Questions: None; Ask Anything\\
Reasoning: Let's think step by step in order to produce the new\_question. We can start by looking at the content of the email. The sender is discussing a lunch meeting and also mentions a few sports teams, including OU and K-State. They also mention that their daughter is a fan of NU and has attended Baylor games. We can ask a question that tests the reader's understanding of the sender's family.\\
New Question: What team is the sender's daughter a fan of?\\
\\
---\\
\end{tcolorbox}

\begin{tcolorbox}[colback=blue!5!white, colframe=blue!75!black, title=Initial QA Generation Prompt (DSPy Optimized) (Part 2)]

Email: Subject: Re: fritzbrinkman resume\\Sender: mike.mcconnell@enron.com\\Recipients: ['robert.jones@enron.com']\\File: mcconnell-m/all\_documents/442.\\=====================================\\Robert,\\\\Thanks,  I've talked with EES - Tom White and I will forward on his resume \\for consideration.  I will also talk with Stan Horton.\\\\Mike\\=====================================\\
Prior Questions: 1.Who did Mike McConnell say he would forward the resume to for consideration, and who else did he plan to talk with?\\
Reasoning: Let's think step by step in order to produce the new\_question. We need to analyze the email and the prior questions to identify a new aspect of the email that hasn't been covered yet. The prior question focused on the actions Mike McConnell would take with the resume. A new question could focus on the context of the email, such as the relationship between the sender and the recipient.\\
New Question: What is the apparent relationship between Mike McConnell and Robert Jones, based on the tone and content of the email about fritzbrinkman's resume?\\
\\
---\\
\\
Email: Subject: Asm Dem Plan\\Sender: jeff.dasovich@enron.com\\Recipients: ['richard.shapiro@enron.com']\\File: dasovich-j/sent/4786.\\=====================================\\Here's the document that I received from the lobbyist today.  It's the same \\document that I distributed internally yesterday.\\\\Best,\\Jeff\\----- Forwarded by Jeff Dasovich/NA/Enron on 05/31/2001 05:26 PM -----\\Delaney Hunter" dhunter@s-k-w.com\\05/30/2001 12:29 PM\\Please respond to dhunter\\ \\ To: Jeff Dasovich (E-mail) jdasovic@enron.com\\ cc: \\ Subject: Asm Dem Plan\\\\\\A little something we got our hands on.....\\ - Assembly Demo Plan.doc\\=====================================\\
Prior Questions: 1.What is the name of the document Jeff Dasovich received from the lobbyist, and who originally forwarded it to him?\\
Reasoning: Let's think step by step in order to produce the new\_question. We can analyze the email to identify the key elements that have not been covered by the prior questions. The prior question focused on the document's name and the person who forwarded it. We can look for other specific details in the email that can be used to create a new question. One detail that stands out is the date and time the email was forwarded by Jeff Dasovich. We can create a question that tests the reader's understanding of this specific detail.\\
New Question: On what date and time did Jeff Dasovich forward the document from the lobbyist to Richard Shapiro?
\end{tcolorbox}

\begin{tcolorbox}[colback=blue!5!white, colframe=blue!75!black, title=Initial QA Generation Prompt (DSPy Optimized) (Part 3)]
---\\
\\
Email: Subject: Service deals\\Sender: holden.salisbury@enron.com\\Recipients: ['portland.shift@enron.com']\\File: guzman-m/all\_documents/1285.\\=====================================\\If you are entering a deal for Harbor, LV Cogen, Willamette, or Saguaro copy \\a deal from the DEAL TEMPLATE worksheet on the Services Sheet.  These deals \\are coded to show our profits.  Feel free to change any of the cells except \\the BUY PRICE and P L that are highlighted in blue.  These cells have \\formulas that should not be changed.  Make sure that you input the correct \\BASIS price as well.\\\\Thanks,\\Holden\\=====================================\\
Prior Questions: None; Ask Anything\\
Reasoning: Let's think step by step in order to produce the new\_question. We need to identify the main topic of the email, which is about entering deals for specific power plants. We also need to consider what information the sender is trying to convey, which is about using a template and following certain guidelines. We can ask a question that tests the reader's understanding of the specific guidelines mentioned in the email. \\
New Question: What two cells in the deal template should not be changed by the user, according to Holden's email about entering deals in the Services Sheet? \\
\\
---\\
\\
Email: \{\textbf{email}\}\\
Prior Questions: \{\textbf{prior questions}\}\\
Reasoning: Let's think step by step in order to
\end{tcolorbox}

\subsection{Email Selection Prompt}
\label{sec:email-selection-prompt}
This prompt is used to measure the specificity of the question.  If Llama3.1 70b is able to select the correct email which corresponds to the question out of a list of 10 emails, then we deem the question to be specific.

\begin{tcolorbox}[colback=blue!5!white, colframe=blue!75!black, title=Email Selection Prompt]
Given a set of emails and a question, select the email that best answers the question.\\
\\
---\\
\\
Follow the following format.\\
\\
Emails: The emails we want to select from\\
Question: The question we want to answer\\
Reasoning: Let's think step by step in order to \$\{produce the selected\_email\}\\
Selected Email: The number corresponding to the email that best answers the question\\
\\
---\\
\\
Emails: \{\textbf{emails}\}\\
Question: \{\textbf{question}\}\\
Reasoning: Let's think step by step in order to
\end{tcolorbox}

\subsection{QA Refinement for Specificity Prompt}
\label{sec:refinement-specificity-prompt}
Here we include both the unoptimized prompt for QA refinement to make questions more specific as well as the DSPy MIPROv2 \cite{opsahlong2024optimizinginstructionsdemonstrationsmultistage} optimized prompt including a rewritten instruction and one bootstrapped fewshot example.  This prompt is used to rewrite questions so that they are more specific and cannot accidentally refer to several different emails (or be answered by several different emails).  This is optimized in the same end-to-end optimization described in \S \ref{sec:initial-qa-generation-prompt}.

\begin{tcolorbox}[colback=blue!5!white, colframe=blue!75!black, title=QA Refinement For Specificity (Unoptimized)]
Given a question and an associated email, alongside a set of unrelated but similar emails, refine the question to be more specific to the EXACT email in question.  You may consider adding details from the email to the question.  Don't significantly change the meaning of the question, just make it more specific to the email in question, and answerable with a single sentence.\\
\\
---\\
\\
Follow the following format.\\
\\
Email: The email we want to refine the question for\\
Question: The question we want to refine\\
Other Emails: The set of other emails which we DO NOT want the question to be about.  Ensure that the refined question has details that do not apply to these emails. \\
Reasoning: Let's think step by step in order to \$\{produce the new question\}. We ...\\
New Question: The brand new question we want to ask about the email\\
\\
---\\
\\
Email: \{\textbf{email}\}\\
Question: \{\textbf{question}\}\\
Other Emails: \{\textbf{other emails}\}\\
Reasoning: Let's think step by step in order to
\end{tcolorbox}

\begin{tcolorbox}[colback=blue!5!white, colframe=blue!75!black, title=QA Refinement For Specificity (DSPy Optimized) (Part 1)]
You are a high-stakes investigator tasked with uncovering the truth behind a series of mysterious emails. You have a question about a specific email, but you need to refine it to ensure it's specific to that exact email and not applicable to other similar emails. The fate of the investigation rests on your ability to craft a precise question. Given the email and a set of unrelated but similar emails, refine the question to make it more specific to the exact email in question, adding details from the email as necessary. The question must be answerable with a single sentence. The entire investigation is counting on you - refine the question carefully.\\
\\
---\\
\\
Follow the following format.\\
\\
Email: The email we want to refine the question for\\
Question: The question we want to refine\\
Other Emails: The set of other emails which we DO NOT want the question to be about.  Ensure that the refined question has details that do not apply to these emails. \\
Reasoning: Let's think step by step in order to \$\{produce the refined\_question\}. We ...\\
Refined Question: The refined question that is more specific to the email in question\\
\\
---\\
\\
Email: Subject: Asm Dem Plan\\Sender: jeff.dasovich@enron.com\\Recipients: ['richard.shapiro@enron.com']\\File: dasovich-j/sent/4786.\\=====================================\\Here's the document that I received from the lobbyist today.  It's the same \\document that I distributed internally yesterday.\\\\Best,\\Jeff\\----- Forwarded by Jeff Dasovich/NA/Enron on 05/31/2001 05:26 PM -----\\Delaney Hunter dhunter@s-k-w.com\\05/30/2001 12:29 PM\\Please respond to dhunter\\ \\ To: Jeff Dasovich (E-mail) jdasovic@enron.com\\ cc: \\ Subject: Asm Dem Plan\\\\\\A little something we got our hands on.....\\ - Assembly Demo Plan.doc\\=====================================\\
Question: On what date and time did Jeff Dasovich forward the email to Richard Shapiro? \\
Other Emails: $\backslash$u00ab1$\backslash$u00bb.test - - - - - original message - - - - - from : dasovich, jeff sent : wednesday, november 21, 2001 12 : 37 pm to : dasovich, jeff subject : test\\$\backslash$u00ab2$\backslash$u00bb.- - - - - original message - - - - - from : dasovich, jeff sent : saturday, october 06, 2001 2 : 44 pm to :'jdasovich @ sbcglobal. net'subject :\\$\backslash$u00ab3$\backslash$u00bb.Here's the document that I received from the lobbyist today.  It's the same  document that I distributed internally yesterday.  Best, Jeff ----- Forwarded by Jeff Dasovich/NA/Enron on 05/31/2001 05:26 PM -----  \"Delaney Hunter\" <dhunter@s-k-w.com> 05/30/2001 12:29 PM Please respond to dhunter    To: \"Jeff Dasovich (E-mail)\" <jdasovic@enron.com>  cc:   Subject: Asm Dem Plan   A little something we got our hands on.....  - Assembly Demo Plan.doc
\end{tcolorbox}

\begin{tcolorbox}[colback=blue!5!white, colframe=blue!75!black, title=QA Refinement For Specificity (DSPy Optimized) (Part 2)] 
$\backslash$u00ab4$\backslash$u00bb.- - - - - forwarded by jeff dasovich / na / enron on 04 / 19 / 2001 10 : 43 am - - - - - jeff dasovich sent by : jeff dasovich 04 / 19 / 2001 10 : 43 am to : psellers @ haas. berkeley. edu cc : subject :\\$\backslash$u00ab5$\backslash$u00bb.reply message : reply from rick shapiro is thank you to jeff. dasovich @ enron. com | | talked to pat. he wants original message : jeff. dasovich @ enron. com | | talked to pat. he wants me to talk by phone rather than come to dc. happy to go if necessary. chk email for leg update. best, jeff\\$\backslash$u00ab6$\backslash$u00bb.reply message : reply from rick shapiro is thank you to jeff. dasovich @ enron. com | | talked to pat. he wants original message : jeff. dasovich @ enron. com | | talked to pat. he wants me to talk by phone rather than come to dc. happy to go if necessary. chk email for leg update. best, jeff\\
Reasoning: Let's think step by step in order to refine the question. We can see that the question is asking for the date and time that Jeff Dasovich forwarded the email to Richard Shapiro. However, the question does not specify which email is being referred to. We can refine the question by adding details from the email that make it more specific. We can see that the email is about the "Asm Dem Plan" and that it was forwarded by Jeff Dasovich on 05/31/2001 05:26 PM. We can also see that the email was originally sent by Delaney Hunter on 05/30/2001 12:29 PM. We can use these details to refine the question.\\
Refined Question: On what date and time did Jeff Dasovich forward the email with the subject "Asm Dem Plan" and attachment "Assembly Demo Plan.doc" to Richard Shapiro?
\\
---\\
\\
Email: \{\textbf{email}\}\\
Question: \{\textbf{question}\}\\
Other Emails: \{\textbf{other emails}\}\\
Reasoning: Let's think step by step in order to
\end{tcolorbox}

\subsection{QA Refinement from Feedback Prompt}
\label{sec:refinement-feedback-prompt}
Here we include both the unoptimized prompt for QA refinement to make questions higher quality as well as the DSPy MIPROv2 \cite{opsahlong2024optimizinginstructionsdemonstrationsmultistage} optimized prompt including a rewritten instruction and two bootstrapped fewshot examples.  This prompt is used in the refinement step to make questions higher quality based on the automatically generated feedback.  This is optimized in the same end-to-end optimization described in \S \ref{sec:initial-qa-generation-prompt}.

\begin{tcolorbox}[colback=blue!5!white, colframe=blue!75!black, title=QA Refinement From Feedback (Unoptimized)]
Given a question and an associated email, alongside feedback on the quality of the question, refine the question to address the feedback.  Don't significantly change the meaning of the question, just incorporate the feedback and keep the question answerable with a single sentence.\\
\\
---\\
\\
Follow the following format.\\
\\
Email: The email we want to refine the question for\\
Question: The question we want to refine\\
Feedback: Feedback on the quality of the question\\
Reasoning: Let's think step by step in order to \$\{produce the new question\}. We ...\\
Refined Question: The refined question that addresses the feedback\\
\\
---\\
\\
Email: \{\textbf{email}\}\\
Question: \{\textbf{question}\}\\
Feedback: \{\textbf{feedback}\}\\
Reasoning: Let's think step by step in order to
\end{tcolorbox}

\begin{tcolorbox}[colback=blue!5!white, colframe=blue!75!black, title=QA Refinement From Feedback (DSPy Optimized) (Part 1)]
You are a high-stakes team lead tasked with refining questions for a critical project for a Fortune 500 company. Your goal is to produce high-quality, objective questions that can be used to train AI models. You have been provided with an email, a question related to the email, and feedback from an expert on the quality of the question. Your task is to refine the question to address the feedback, without significantly changing its meaning, and ensuring it remains answerable with a single sentence. The quality of your refined question will directly impact the success of the project and the company's reputation. Refine the question carefully, and provide your reasoning for the changes you make.\\
\\
---\\
\\
Follow the following format.\\
\\
Email: The email we want to refine the question for\\
Feedback: Feedback on the quality of the question\\
Reasoning: Let's think step by step in order to \$\{produce the new question\}. We ...\\
Refined Question: The refined question that addresses the feedback\\
\\
---\\
\\
Email: Subject: FW: List of authorized traders\\Sender: alterson\_thomas@jpmorgan.com\\Recipients: ['sara.shackleton@enron.com']\\File: shackleton-s/all\_documents/11472.\\=====================================\\Have you been able to take a look at this?\\\\I hate to be a pest, but we are in the midst of an internal audit right now \\and\\we are being pressured for this list.\\\\Thank you for your time.\\Tom Alterson\\\\\\Thomas P Alterson\\06/05/2001 12:18 PM\\\\To:   Sara.Shackleton@enron.com\\cc:\\Subject:  List of authorized traders  (Document link not converted)\\\\Sara:\\\\Would you please send me a list of authorized traders for Enron NA Corp?  We \\are\\updating our files and noticed that we are missing an ATL for Enron.\\\\*ATL form:\\(See attached file: ATL Enron NA Corp.doc)\\\\Thank you,\\Tom Alterson\\ - ATL Enron NA Corp.doc\\=====================================
\end{tcolorbox}

\begin{tcolorbox}[colback=blue!5!white, colframe=blue!75!black, title=QA Refinement From Feedback (DSPy Optimized) (Part 2)]
Question: What is the reason given by Tom Alterson for needing the list of authorized traders for Enron NA Corp?\\
Feedback: Question is not objective.  Different annotators answer the same question differently given the same email as context.  Could benefit from more clarity.\\
Reasoning: Let's think step by step in order to refine the question. We need to make the question more objective and clear. The reason given by Tom Alterson for needing the list of authorized traders for Enron NA Corp is mentioned in the email as "we are updating our files and noticed that we are missing an ATL for Enron." However, this reason is not explicitly stated as the main reason, but rather as a secondary reason. The primary reason is mentioned in the forwarded email as "we are in the midst of an internal audit right now and we are being pressured for this list." To make the question more objective, we can rephrase it to focus on the primary reason mentioned in the email.\\
Refined Question: What is the primary reason given by Tom Alterson for needing the list of authorized traders for Enron NA Corp during the internal audit?\\
\\
---\\
\\
Email: Subject: FW: List of authorized traders\\Sender: alterson\_thomas@jpmorgan.com\\Recipients: ['sara.shackleton@enron.com']\\File: shackleton-s/all\_documents/11472.\\=====================================\\Have you been able to take a look at this?\\\\I hate to be a pest, but we are in the midst of an internal audit right now \\and\\we are being pressured for this list.\\\\Thank you for your time.\\Tom Alterson\\\\\\Thomas P Alterson\\06/05/2001 12:18 PM\\\\To:   Sara.Shackleton@enron.com\\cc:\\Subject:  List of authorized traders  (Document link not converted)\\\\Sara:\\\\Would you please send me a list of authorized traders for Enron NA Corp?  We \\are\\updating our files and noticed that we are missing an ATL for Enron.\\\\*ATL form:\\(See attached file: ATL Enron NA Corp.doc)\\\\Thank you,\\Tom Alterson\\ - ATL Enron NA Corp.doc\\=====================================\\
Question: What is the primary reason given by Tom Alterson for needing the list of authorized traders for Enron NA Corp during the internal audit?
\end{tcolorbox}

\begin{tcolorbox}[colback=blue!5!white, colframe=blue!75!black, title=QA Refinement From Feedback (DSPy Optimized) (Part 3)]
Feedback: Question is not objective.  Different annotators answer the same question differently given the same email as context.  Could benefit from more clarity.\\
Reasoning: Let's think step by step in order to refine the question. We need to make the question more objective and clear. The primary reason given by Tom Alterson for needing the list of authorized traders for Enron NA Corp is mentioned in the email as "we are in the midst of an internal audit right now and we are being pressured for this list." This reason is related to the internal audit, but the question could be more specific. We can refine the question to ask about the specific event or situation that is causing the pressure for the list.\\
Refined Question: What event or situation is causing pressure for the list of authorized traders for Enron NA Corp, according to Tom Alterson?\\
\\
---\\
\\
Email: \{\textbf{email}\}\\
Question: \{\textbf{question}\}\\
Feedback: \{\textbf{feedback}\}\\
Reasoning: Let's think step by step in order to
\end{tcolorbox}

\subsection{Question Answering Prompts}
\label{sec:question-answering-prompts}
These prompts are used to both answer the question given the context of an email or to produce an answer to the question with no grounding.  Forcing the LLM to answer the question without grounding is to ensure that the questions are not too easy to guess or and are not memorized by popular LLMs.

\begin{tcolorbox}[colback=blue!5!white, colframe=blue!75!black, title=QA With Email Prompt]
Given an email and a question about that email, write the answer to that question in a single sentence.\\
\\
---\\
\\
Follow the following format.\\
\\
Email: The email we want to answer a question about\\
Question: The question we want to answer about the email\\
Reasoning: Let's think step by step in order to \$\{produce the answer\}\\
Answer: The answer to the question\\
\\
---\\
\\
Email: \{\textbf{email}\}\\
Question: \{\textbf{question}\}\\
Reasoning: Let's think step by step in order to
\end{tcolorbox}

\begin{tcolorbox}[colback=blue!5!white, colframe=blue!75!black, title=QA Without Email Prompt]
Given a question, write an answer to that question in a single sentence.\\
\\
---\\
\\
Follow the following format.\\
\\
Question: The question we want to answer\\
Reasoning: Let's think step by step in order to \$\{produce the answer\}\\
Answer: The answer to the question\\
\\
---\\
\\
Question: \{\textbf{question}\}\\
Reasoning: Let's think step by step in order to
\end{tcolorbox}

\subsection{LLM as a Judge Prompt}
\label{sec:llm-as-a-judge-prompt}
This prompt was used for our LLM as a judge to determine whether or not two answers were the same or different.  The LLM as a judge was grounded in the document which helped it determine if additional details in a particular answer were a hallucination or grounded in factual information.  We sampled 100 instances where the LLM as a judge deemed answers to match, and 100 instances where the LLM as a judge deemed answers as not matches.  An author manually labelled these assessments and we determined the LLM as a judge to have 0.98 F1-score (only differing in 2 judgements in both cases with the human judge).  This gave us high confidence in using our LLM judge thoughout our evaluation.  It is important to note that we are not using an LLM as a judge to make subjective judgment calls here, but rather to determine if two open-ended answers match or not.  This explains the high accuracy even when LLM as a judge can be unreliable \cite{wei2024systematicevaluationllmasajudgellm}.

\begin{tcolorbox}[colback=blue!5!white, colframe=blue!75!black, title=LLM as a Judge]
Given an email, a question about that email, a gold answer to that question, and a student's potentially correct or incorrect response, judge whether the answer matches the gold answer.\\
\\
---\\
\\
Follow the following format.\\
\\
Email: The email we want to judge the answer to\\
Question: The question we want to judge the answer to\\
Correct Answer: The correct answer to the question\\
Student Answer: The student answer we want to judge\\
Reasoning: Let's think step by step in order to \$\{produce the correct\}\\
Answer: Whether the answer is correct\\
\\
---\\
\\
Email: \{\textbf{email}\}\\
Question: \{\textbf{question}\}\\
Correct Answer: \{\textbf{correct answer}\}\\
Student Answer: \{\textbf{student answer}\}\\
Reasoning: Let's think step by step in order to
\end{tcolorbox}

\subsection{Rule Based Quality Evaluation Prompt}
\label{sec:quality-evaluation-prompts}
These prompts are used to both answer the question given the context of an email or to produce an answer to the question with no grounding.  Forcing the LLM to answer the question without grounding is to ensure that the questions are not too easy to guess or and are not memorized by popular LLMs.

\begin{tcolorbox}[colback=blue!5!white, colframe=blue!75!black, title=Rule Based Quality Evaluation]
Given a particular email, a question about the email, and the answer to that question, judge the quality of the question.  The question should follow the following criteria:\\
1. The question should be specific to the factual contents of this email.  In other words, if you were given 100 emails, would you know that the question was about this email and not another one?  Phrases such as "the email" or "the message" are not specific enough unless grounded by further context.  Just the sender alone may not be enough context, but the sender and some detail about the email might be.\\
2. The question should focus on the main contents of the message, not on the formatting, the sender, or recipients.  It is okay to use the sender and recipients as context, but the question should not be about them.  It is okay to ask about things like the cell phone numbers or contact details of the people in the email since this is a realistic question.\\
3. The question should be objective and answerable with a single sentence.  It should not be a matter of opinion or require any interpretation.\\
4. The question should be realistic to what a person might ask about an email they received, especially in the context of working in a professional setting and recalling important details.\\
5. The question should NOT require any external knowledge beyond the contents of the email itself.\\
6. The question should NOT include counting, math, or any other operations.  For example "How many times did the sender mention the word 'important'?" or "how many recipients were there?" are not allowed.  It is okay for the question to ask about a number in the email such as a percentage, but it should not involve actually counting or calculating.\\
Given these rules, rate the question as either "good" or "bad".\\
\\
---\\
\\
Follow the following format.\\
\\
Email: The email we want to answer a question about\\
Question: The question we want to answer about the email\\
Answer: The answer to the question\\
Reasoning: Let's think step by step in order to \$\{produce the quality\}\\
Quality: Whether the question is good (true) or bad (false)\\
\\
---\\
\\
Email: \{\textbf{email}\}\\
Question: \{\textbf{question}\}\\
Answer: \{\textbf{answer}\}\\
Reasoning: Let's think step by step in order to
\end{tcolorbox}

\end{document}